# A Machine Learning Approach to Detect Suicidal Ideation in US Veterans Based on Acoustic and Linguistic Features of Speech

**Vaibhav Sourirajan**
Winston Churchill High School, Potomac, MD
Summer Internship (2019-2020) at Georgetown University
*vs625@georgetown.edu*

**Research Mentors**
Anas Belouali MS[1], Mary Ann Dutton PhD[2], Matthew Reinhard PsyD[2], Jyotishman Pathak PhD[3]
[1]Georgetown University, Washington DC; [2]Medstar Georgetown University Hospital, Washington DC; [3]US Department of Veterans Affairs, Washington DC; [4]Weill Cornell Medical College, NY

**Abstract**

*Preventing Veteran suicide is a national priority. The US Department of Veterans Affairs (VA) collects, analyzes, and publishes data to inform suicide prevention strategies. Current approaches for detecting suicidal ideation mostly rely on patient self-report which are inadequate and time-consuming. In this research study, our goal was to automate suicidal ideation detection from acoustic and linguistic features of an individual's speech using machine learning (ML) algorithms. Using voice data collected from Veterans enrolled in a large interventional study on Gulf War Illness at the Washington DC VA Medical Center, we conducted an evaluation of the performance of different ML approaches in achieving our objective. By fitting both classical ML and deep learning models to the dataset, we identified the algorithms that were most effective for each feature set. Among classical machine learning algorithms, the Support Vector Machine (SVM) trained on acoustic features performed best in classifying suicidal Veterans. Among deep learning methods, the Convolutional Neural Network (CNN) trained on linguistic features performed best. Our study showed that speech analysis in a machine learning pipeline is a promising approach for detecting suicidality among Veterans.*

## I. Introduction

Suicide prevention continues to remain a challenging clinical issue, especially among Veterans. The United States Department of Veteran Affairs (VA) collects, analyzes, and publishes data to inform prevention strategies and programs. According to the 2019 National Veteran Suicide Prevention Annual Report[1], Veterans are 1.5 times more likely to die by suicide compared to non-Veteran adults with 17 Veterans dying per day due to this cause. From 2005 to 2017, the suicide rate among US adults increased by 22.4% while the suicide rate among Veteran adults rose 49%.

Currently screening for suicidality is a complex dynamic, which heavily relies on the patient disclosing their intent (self-report), plans, or behavior for self-harm to their healthcare provider. Healthcare providers utilize one of several self report screening tools such as Suicidal Ideation Questionnaire (SIQ)[2] and Post Traumatic Stress Disorder Checklist (PCL)[3]. However, due to associated stigma, especially amongst the Veterans, patient self-report data is often considered as biased, inadequate, and untimely[4]. As a result, research into objective markers for clinical assessments is crucial in identifying and preventing suicide.

Latest advances in digital technologies enable collection of novel data streams from patients for suicide prevention research. One such novel data stream is longitudinal speech, an information-rich signal and measurable form of behavior that can be collected outside of clinical care. This enables real-time and context-aware monitoring of an individual's mental state. Prior studies have demonstrated the feasibility of speech analysis and machine learning in detecting mental health disorders[5,6]. In 2018, researchers utilized a recurrent neural network implementation to classify depressed and healthy speech among patients undergoing depression screening[7]. The authors in this study predicted depression using sequentially collected audio interviews from patients with an accuracy (binary F1 score[8]) of 0.43-0.77. Another study examining mania and depression in bipolar disorder aimed to investigate voice features from phone calls as objective markers of affective states, classifying patients with an Area Under the Receiver-Operating Characteristic Curve[9] (AUC) of 0.89[6].

Studies examining voice data from suicidal patients date back to 1992, describing suicidal voices as hollow, toneless, monotonous with mechanical and repetitive phrasing[10–12]. A research study conducted in 2013 examined speech characteristics in suicidal adolescents, revealing statistically significant differences in speech patterns compared to non suicidal adolescents. Specifically, the research team identified glottal features in voice - physiological characteristics of the vocal folds when an individual is under different emotional modes - that strongly differentiated between the two groups[13]. Recent work investigated the effectiveness of acoustic and linguistic features of speech in discriminating between the conversation of suicidal and non suicidal individuals. Classifying 379 patients into three categories (suicidal, mentally ill but non-suicidal, and controls), the ML algorithms performed with an accuracy of up to 0.85[14].

Our goal was to investigate an ML-based approach that uses features of speech to detect suicidal ideation in US Veterans. We hypothesized that speech in suicidal Veterans has distinctive acoustic and linguistic features that could detect suicidal ideation. We investigated these features in 515 narrative audios collected longitudinally from 94 Veterans in a naturalistic setting using a mobile application that was previously developed by the VA for data collection purposes. Tablets with the mobile app installed were provided to Veterans for recording their speech on a regular basis, for example, through sleep diaries. To our knowledge, this is the first study to investigate suicidal ideation in US Veterans using longitudinally recorded audios in everyday life settings.

## II. Methodology

### A. Study Data

Data for this research was obtained as part of a large intervention study on Gulf War Illness at the Washington DC VA Medical Center. 149 Veterans meeting the Center for Disease Control's criteria for Gulf War Illness[15] were recruited to the study. 515 audio recordings from 94 Veterans were collected via an Android smartphone app. An Android tablet (Samsung Galaxy Tab 4) with the mobile app installed was provided to each Veteran to enable study participation from their natural setting such as their homes. At each time-point - week 0, week 4, week 8, 3 months, 6 months, 1 year - participants received email notifications or text messages, based on their choice, and were prompted to complete multiple psychiatric assessments. Veterans responded via audio recordings to open-ended prompts about their general health over the past weeks/months. Each Veteran who provided audio recordings also completed a Patient Health Questionnaire[16] (PHQ-9) administered as part of the health questionnaire battery. Item-9 of the PHQ-9 is commonly used in research to screen for suicidality and has been validated to be predictive of suicide in both the general population and in US Veterans[17,18]. The key question in PHQ-9 used in our study was, "Over the last two weeks, how often have you been bothered by thoughts that you would be better off dead or of hurting yourself in some way?" Response options are "not at all", "several days", "more than half the days", or "nearly every day". We considered a subject as suicidal at the time of recording, if they answered with any option other than "not at all". Veteran's responses to this question were used to label the audios used to train the ML models into two classes: suicidal and non-suicidal cases.

### B. Study pipeline

**Figure 1** outlines the pipeline for analysis of the acoustic and linguistic features and their preparation for Classical ML and Deep Learning methods. Briefly, the audio recordings were transcribed using Google Speech-to-Text API[19], and acoustic features were extracted using pyAudioAnalysis[20]. Linguistic features, represented by word embeddings, were generated from the transcribed text corpus using the GENSIM library[21]. Utilizing the Sci-kit Learn[22] and Keras[23] libraries we implemented classical machine learning (Logistic Regression (LR), Random Forest (RF), Support Vector Machine (SVM)) and deep learning algorithms (Artificial Neural Network (ANN), Convolutional Neural Network (CNN)), respectively, on the acoustic and linguistic features separately. All models were trained on the PHQ-9 suicide ideation label with class weights. The prediction performance of each model was evaluated using a 5-Fold Cross Validation (CV) measuring sensitivity, specificity, and area under the receiver-operating curve (AUC).

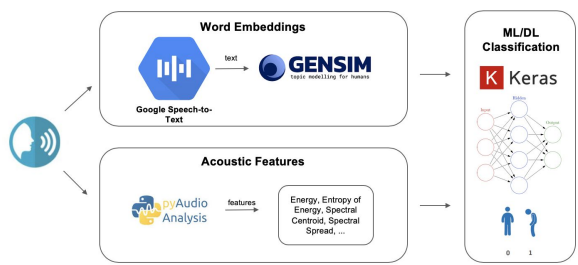

**Figure 1: Machine Learning Pipeline**

The individual pre-processing, exploratory analysis and machine learning methods used in the study are described in sections below.

### C. Pre-processing Methods
**Keras Tokenizer**
The Keras Tokenizer converts the transcribed text into tokenized vectors and prepares the data for deep learning methods. Each message was represented as a tokenized vector based on word frequency: the more frequent the word, the lower its frequency rank. While being fitted on the text corpus, the internal vocabulary was dynamically updated by frequency.

The text sequences were then transformed into sequences of integers, with each word replaced by its corresponding frequency index. All vectors were then padded to the maximum vector length, which in this case represented the longest message that contained 2249 words.

**Word Embeddings**
Word embeddings are numerical vector representations of text. Through word embeddings, a word can be mapped in an n-dimensional vector space, and words with identical meanings have similar vector representations. In this research, we applied the Gensim library to generate word embeddings on the transcribed text, representing each word in the corpus with a 100 dimension vector. The embedding sequences were then used to create an embedding matrix with the tokenized and padded vectors. This embedding matrix was inputted as a parameter in the embedding layer of the Linguistic Feature Neural Networks.

### D. Exploratory analysis
**Scattertext**
We performed linguistic analysis using Scattertext[24] on the transcribed text to examine differences in vocabulary between suicidal Veterans and non-suicidal Veterans. The tool utilizes a scaled f-score, which accounts for category-specific precision and term frequency. As a result, words commonly used by both classes will have scaled f-scores that are characteristic of one class over the other. To have a more accurate representation of class-specific vocabulary, we excluded all stopwords from the corpus. These include words such as the, a, me, is, and in.

**Acoustic Analysis**
A total of 250 acoustic features were extracted from each of the 515 recordings. pyAudioAnalysis[20], an audio signal analysis python library, was used to extract short-term feature sequences using a frame size of 50 milliseconds and a frame step of 25 milliseconds (50% overlap). The pyAudioAnalysis features include: zero crossing rate, energy and entropy of energy, chroma vector and deviation, spectral features composed of centroid, spread, entropy, flux, rolloff and Mel-Frequency Cepstral Coefficients (MFCC) among others. To compare suicidal and non-suicidal speech, these features were investigated by checking their statistical significance using a chi-square test. All raw p-values (p-raw) were adjusted for multiple testing using the Bonferroni correction where p-adj= p-raw x n, where n is the number of independent tests. Statistical significance was defined as padj<0.05. Acoustic features that were significant were further analyzed and discussed with psychiatrists on the research team.

# E. Machine learning approaches
**Random Forest Classifier**
Random Forest Classifiers function by creating several individual decision trees which operate as an ensemble. Each decision tree trains on a subsection of the dataset. The classification prediction is the majority vote of the decision trees. This occurs by taking the average of the estimators for a given sample $x^i$ with the following function:

$$f(x) = \frac{1}{B} \sum_{b=1}^{B} f_b(x)$$

In this equation, the function $f$ represents the class that sample x belongs to, B equals the total number of estimators, and $f_b$ represents the specific class that estimator $b$ predicts sample $x^i$ belongs to. In our implementation of the Random Forest Classifier, we used 5 estimators.

**Support Vector Machine**
Support Vector Machines work by finding a hyperplane vector that divides the two classes with maximal margin. When there is no clear distinction between the two classes, the data is mapped into higher and higher dimensions until a hyperplane can effectively create a separation. This concept is referred to as kernelling.

**Logistic Regression**
The Logistic Regression algorithm uses an iterative optimization algorithm to calculate the parameters of the model. This allows the algorithm to fit an S-shaped curve to the dataset, known as the sigmoid function. This function is defined as the following:

$$sigmoid(x) = \frac{1}{1+e^{-x}}$$

Each test example is then given a probability from 0 to 1 and if the probability exceeds 0.5, the model classifies the example as 1. In our implementation, a probability greater than 0.5 would classify a suicidal patient.

**Artificial Neural Network**
Artificial Neural Networks (ANNs) work by funneling input data through a series of weighted layers. This technique is directly modeled after the connection between synapses in the human brain. While being fit on the training data, the ANN internalizes patterns between the input data and its corresponding class. This gives ANNs their predictive abilities.

We implemented the ANN algorithm on both feature sets. In the Acoustic Features ANN, we included 3 hidden Dense Layers of 128, 64, and 32 nodes respectively. All three hidden layers were activated by the ReLU function defined as the following:

$$ReLU(x) = max(0, x)$$

In the Linguistic Features ANN, we included an Embedding Layer, which embeds the tokenized vectors and the generated word embeddings. Both ANN algorithms' 1-node output layers were activated by the sigmoid function.

We trained both ANNs for 250 epochs using the Adam optimizer and the binary cross entropy loss function defined as the following:

$$H_p(q) = -\frac{1}{N} \sum_{i=1}^{N} y_i \cdot log(p(y_i)) + (1-y_i) \cdot log(1-p(y_i))$$

**Convolutional Neural Network**
Convolutional Neural Networks (CNNs) are a subset of ANNs, but include a key difference: Convolutional Layers. These layers contain a set of filters, each of which learns a local characteristic of the input (subset of features, particular phrase or group of words). These filters are convolved with the input volume to create a feature map that prioritizes repeated overlaps. Each value in the feature map is then passed through the ReLU activation function.

CNNs are especially beneficial, as opposed to standard ANNs, because they take into account the order and sequences within the data. Additionally, the creation of a feature/activation map allows CNNs to learn the most important features in the dataset, and thus make better classification predictions

The Convolutional Layer in both CNNs included 250 filters. Much like the ANNs, we trained both CNNs for 250 epochs using the Adam optimizer and the binary cross entropy loss function.

**Measurement Metrics and Validation Technique**
This data was observed to have a large class imbalance with one suicidal recording for every six non-suicidal recordings. As a result, accuracy was not a proper indicator of model performance. Consequently, we measured sensitivity, specificity, and AUC, all of which account for imbalance and indicate a model's capability to distinguish between classes. We reported all metrics jointly. To evaluate model performance, we performed a K-Fold Cross Validation (CV) with 5 folds. During each iteration, we train on four of the five folds--80% of the dataset-- and test on the remaining fold--20% of the dataset. In addition, we partitioned the data using a Stratified Split. This ensured that each fold maintained the 1:6 class ratio of the original dataset. The values for sensitivity, specificity, and AUC reflect an average of the five iterations.

## III. Results

**Veteran demographics and their recordings**
During the time period between May 2016 and January 2020, as part of the Gulf War Illness Veterans Study conducted by the US Department of Veterans Affairs, 149 Veterans were recruited. 515 narrative audios from 94 Veterans from this study were used for this research project. Based on the responses of Veterans to questions in the PHQ-9 questionnaire, 73 audios were classified as suicidal and 442 were non-suicidal. The average age of this group was 52.4 years (std= 9.4) and the majority of participants were male Veterans (79%).

**Exploratory Analysis Results**
From acoustic analysis, there were no major differences between suicidal and non-suicidal recordings in audio length, loudness, or duration of pauses. Suicidal recordings were mainly different from non-suicidals in terms of energy. Suicidal speech had a flatter, less bursty, and less animated voice. Suicidal audios were dull and more monotonous voices and had breathier voices.

Linguistic analysis using Scattertext (**Figure 2**) outlined the top words used by both suicidal and non-suicidal Veterans. The tool analyzed over 40 thousand words from the text corpus to assign a scaled F-Score to each word. After examining the scaled f-scores, words and phrases strongly associated with suicidal ideation (red dots) were 'chronic', 'problems sleeping', 'severe' and 'migraine', while words and phrases strongly associated with non-suicidality (blue dots) were 'okay', 'right', 'helping' and 'appreciate'.

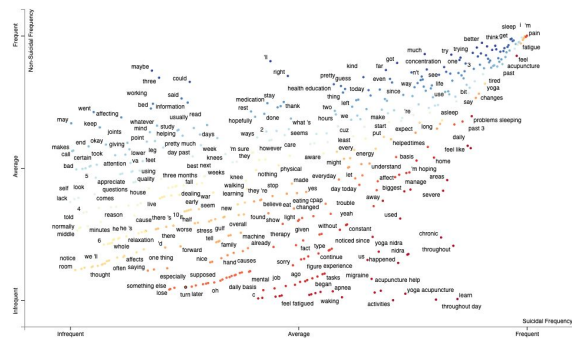

Figure 2: Scattertext visualization of words associated with suicidal and non-suicidal Veterans

**Machine Learning Results**
We tested the hypothesis that speech has a range of distinctive acoustic and linguistic features that could identify suicidal ideation in Veterans and therefore assessed these features in ML approaches separately. Among classical machine learning algorithms, the SVM trained on acoustic features performed best with a sensitivity of 0.63, specificity of 0.64 and area under the receiver operating curve of 0.64. Among deep learning methods, the CNN trained on linguistic features performed best with a sensitivity of 0.64, specificity of 0.65 and area under the receiver operating curve of 0.65. The complete results are shown in **Table 1**. The AUC curves for acoustic and linguistic features are shown in **Figure 3.**

| Table 1: Binary Classification Results Based on Acoustic and Linguistic Features of Speech | | | | |
|---|---|---|---|---|
| Feature Set | Model | Sensitivity | Specificity | AUC |
| Acoustic | RF | 0.63 | 0.62 | 0.63 |
| | **SVM** | **0.63** | **0.64** | **0.64** |
| | LR | 0.51 | 0.55 | 0.53 |
| | ANN | 0.67 | 0.60 | 0.63 |
| | CNN | 0.55 | 0.62 | 0.58 |
| Linguistic | RF | 0.53 | 0.52 | 0.53 |
| | SVM | 0.51 | 0.51 | 0.51 |
| | LR | 0.52 | 0.61 | 0.56 |
| | ANN | 0.52 | 0.56 | 0.54 |
| | **CNN** | **0.64** | **0.65** | **0.65** |

## IV. Discussion

Our findings indicate that ML-based speech analysis is a promising approach for detecting suicidality. The choice to examine distinctive acoustic and linguistic features was important to optimize the performance of the classifiers on each type of data and provide realistic estimations on suicidality. Exploratory analysis indicated that acoustic features representing pitch, flatness, energy, breathiness, and tenseness in voice showed the strongest difference between the suicidal and non-suicidal groups. We found that suicidal speech had less energy, was flat, less animated and involved breathiness compared to non-suicidal speech. Investigation of text features demonstrated positive sentiment among Veterans who were non-suicidal compared to words of negative emotions used by suicidal Veterans. We also observed trends indicating more superlative adverbs, possessive pronouns, and personal nouns in suicidal speech. Overall, we found that suicidal Veterans spoke with certainty (e.g. certain, certainly) discussing topics such as chronic pain (pills, knees) or sleep problems (CPAP machine) when describing their general health in the past weeks and months. Conversely, non-suicidal Veterans used action verbs and words indicating improvements (e.g. function, improve, trying, find, noticed, aware).

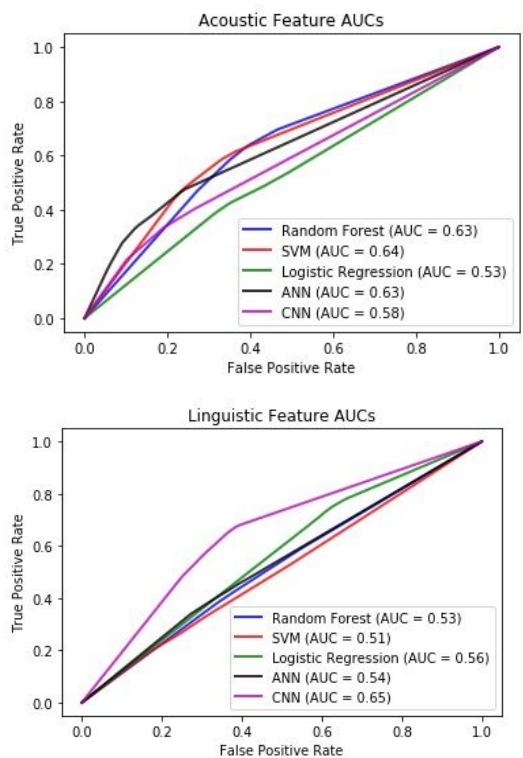

Figure 3: Area Under the Receiver Operating Characteristic Curve (AUC) for Acoustic features and Linguistic features

The primary focus on this research work was building and testing classification models for suicidality. The results show that acoustic-based models performed better (AUC=0.60) than models based on linguistic features alone (AUC=0.56). Previous research on suicidality and speech in other risk groups reported varying performance metrics and relied on smaller sample size [13,14,25]. While our performance was moderate, in future work, we plan to combine acoustic and linguistic features using ensemble feature selection or stacked model implementation to further optimize the classifiers. We also plan to implement sampling techniques such as Synthetic Minority Oversampling Technique (SMOTE)[26] and near-miss undersampling[27]. External validation outside of the VA is also needed to demonstrate broad utility of this approach. EHR or biometric data from patients may be useful in improving the classifiers. Our research study had a few limitations. We relied on a self report to indicate whether a subject was suicidal or not at the time of the recording. Hence, it is possible that some of the recordings were mislabeled if a participant did not accurately disclose their suicidal state either due to stigma or incorrect reporting on the PHQ-9 questionnaires. Furthermore, the audios ranged from a few seconds to several minutes long and were recorded in various everyday life settings (e.g. home, park, work place) which could have introduced background noise and quality issues.

**Public Health Relevance**

This research can expand beyond just a single clinical issue and subset of the population, suicidal ideation and US Veterans respectively. For example, preventing suicide among youth is a growing national concern. In the United States alone, youth suicide rates rose 56% from 2007-2017 and is currently the second leading cause of death for ages 10-24[28]. The research presented here has the potential to address critical public health concerns, not just among Veterans, but also among other demographics including teenagers, young adults, and trauma victims.

**Broader Impact**

Automated mental health assessment and risk prediction using routine speech data presented here have downstream applications in developing emotionally-intelligent and empathetic companion robots that will detect mental health illnesses early, perform risk stratification, and make appropriate treatment recommendations. Such emotionally intelligent human companion robots are being envisioned and developed across the globe. Examples include:
- NASA exploring emotional robots to accompany astronauts during the long Mars journey
- *Huggable* companion robot at Boston Children's Hospital to ease hospital stress among children
- *Robin*, a companion robot in Armenian hospitals to comfort children and young adults during emergency procedures

**Ethical considerations**

The AI methods and their potential uses in emotionally-intelligent robot companions presented here raise many ethical considerations regarding what they should and should not do. Robots that could think and feel like humans could potentially cause harm. There are many active ethical debates such as privacy and surveillance[29] (e.g. in-home surveillance cameras), behavior modification[30] (e.g. use of AI in targeted marketing by companies to maximize profit), and bias in predictive systems[31] (e.g. predictive policing such as in the movie 2002 Minority report). In a healthcare setting, this might result in de-humanized care[32]. By no means do we have answers to all these questions relating to ethical implications of AI. These important issues are being considered by government agencies around the world, which are working to develop a "good AI society"[33]. Additionally, academic AI research centers[34–36] are conducting research on ethical implications of AI and scientific journals are dedicating entire issues to tackle the ethical dilemmas that arise from use of AI technology in healthcare settings[37]. While this was not the primary focus of our research study, we wanted to acknowledge the importance of research and development in the area of ethical considerations in AI to help make many of the innovations discussed here possible in the future.

## V. Conclusions

We showed that speech analysis in a machine learning pipeline is a promising approach in detecting suicidal ideation among US Veterans. This work is applicable to other mental health domains of large public health relevance including anxiety, depression, and homicide/suicide. Overall, our work demonstrates the feasibility of automated approaches that use acoustic and linguistic features of speech in detecting mental health disorders. In the long term, this work can form the basis of emotionally intelligent human companion robots for early detection and timely intervention of mental health disorders

## Acknowledgements
Author VS thanks his research mentors Anas Balouali, Data Scientist at Georgetown University, Dr. Mary Ann Dutton, Psychologist at Medstar Georgetown University Hospital and Dr. Reinhard, Psychiatrist at the US Department of Veterans Affairs, and Dr. Pathak, Computer Scientist at Weill Cornell Medical College for their guidance, timely input and encouragement throughout this research study.

This project was funded partly by NIH NCATS Georgetown-Howard Universities Center for Clinical and Translational Science (GHUCCTS) (UL1-TR001409) and partly with Federal funds (5 I01 CX000801 02) from the U.S. Department of Veterans Affairs Medical Center.